\documentclass{article}

\usepackage{lineno}
\usepackage{arxiv}
\usepackage{tikz}
\usepackage{pgfplots}
\usepackage{graphicx}
\usepackage{algorithm}
\usepackage{algpseudocode}
\usepackage{subcaption}
\usepackage{float}
\usepackage[utf8]{inputenc} 
\usepackage[T1]{fontenc}    
\usepackage{hyperref}       
\usepackage{url}            
\usepackage{booktabs}       
\usepackage{amsfonts}       
\usepackage{nicefrac}       
\usepackage{microtype}      
\usepackage{lipsum}
\usepackage{graphicx}
\graphicspath{ {./images/} }
\usepackage{xcolor} %
\newcommand{\rev}[1]{{\color{black}#1}}

\title{Dynamic Disruption Resilience in Intermodal Transport Networks: Integrating Flow Weighting and Centrality Measures}

\author{
 Aliza Sharmin \\
  Department of Industrial and Systems Engineering\\
  University of Tennessee\\
  Knoxville, TN 37996\\
  \texttt{asharmin@vols.utk.edu} \\
   \And
 Bharat Sharma \\
  Computational Sciences and Engineering Division\\
  Oak Ridge National Laboratory\\
  Oak Ridge, TN 37830\\
  \texttt{sharmabd@ornl.gov} \\
  \And
 Mustafa Can Camur \\
  Department of Industrial and Systems Engineering\\
  University of Tennessee\\
  Knoxville, TN 37996\\
  \texttt{mcamur@utk.edu} \\
  \And
 Olufemi A. Omitaomu \\
  Computational Sciences and Engineering Division\\
  Oak Ridge National Laboratory\\
  Oak Ridge, TN 37830\\
  \texttt{omitaomuoa@ornl.gov} \\
  \And
 Xueping Li \\
  Department of Industrial and Systems Engineering\\
  University of Tennessee\\
  Knoxville, TN 37996\\
  \texttt{xueping.li@utk.edu} \\
}

\begin{document}
\maketitle
\begin{abstract}
Resilient intermodal freight networks are vital for sustaining supply chains amid increasing threats from natural hazards and cyberattacks. While transportation resilience has been widely studied, understanding how random and targeted disruptions affect both structural connectivity and functional performance remains a key challenge. To address this, our study evaluates the robustness of the U.S. intermodal freight network, comprising rail and water modes, using a simulation-based framework that integrates graph-theoretic metrics with flow-weighted centrality measures.
We examine disruption scenarios including random failures as well as targeted node and edge removals based on static and dynamically updated degree and betweenness centrality. To reflect more realistic conditions, we also consider flow-weighted degree centralities and partial node degradation. Two resilience indicators are used: the size of the giant connected component (GCC) to measure structural connectivity, and flow-weighted network efficiency (NE) to assess freight mobility under disruption. \rev{Results show that progressively degrading nodes ranked by Weighted Degree Centrality to 60\% of their original functionality causes a sharper decline in normalized NE, for up to approximately 45 affected nodes, than complete failure (100\% loss of functionality) applied to nodes targeted by weighted betweenness centrality or selected at random.} This highlights how partial degradation of high-tonnage hubs can produce disproportionately large functional losses. The findings emphasize the need for resilience strategies that go beyond network topology to incorporate freight flow dynamics. 

\end{abstract}

\keywords{ Intermodal Freight Networks \and Network Resilience \and Flow-Weighted Centrality \and Partial Node Degradation \and Network Efficiency \and Transportation Infrastructure \and Graph Theory}

\section{Introduction}

In an era of growing global trade and increasingly complex supply chains, the reliability of freight transportation systems is paramount \cite{rodrigue2020geography}. Intermodal transportation, defined as the movement of goods using two or more modes of transport (e.g., rail, waterway, road) within a single journey, has emerged as a key strategy for achieving efficiency, cost-effectiveness, and environmental sustainability \cite{crainic2007intermodal}. By leveraging the strengths of each transport mode, intermodal systems optimize long-haul freight movement while reducing congestion and emissions \cite{bauer2010minimizing}. This has led to a significant increase in the adoption of intermodal logistics across the United States (U.S.) and internationally, driven by both public policy and private-sector initiatives \cite{foxx2017beyond}.

However, the growing reliance on intermodal infrastructure also introduces heightened complexity and vulnerability. The interdependencies between distinct modal subsystems such as inland waterways, rail lines, and highway connectors can amplify the impact of localized disruptions, triggering cascading failures across the broader freight network \cite{chen2012resilience}. Disruptions stemming from extreme weather events, cyberattacks, infrastructure aging, or targeted attack\rev{s} can significantly degrade system performance \cite{markolf2019transportation, mattsson2015vulnerability, sullivan2010identifying,warner2017}. In this context, the resilience of intermodal networks \rev{ defined as their ability to absorb shocks,} maintain functionality, and recover swiftly, becomes a critical attribute for sustaining economic stability and supply chain continuity.

Despite sustained attention to transportation resilience \cite{ganin2017resilience, freckleton2012evaluation, serulle2011resiliency}, critical gaps remain in understanding how intermodal freight networks, particularly  in the U.S., respond to disruptions targeting high-impact nodes or edges, \rev{where nodes correspond to system components such as terminals, ports, or intersections, and edges represent the physical or operational links (e.g., rail lines, waterways, or transfer connections) that enable movement between nodes.} Existing robustness assessments often overlook the interplay between network topology \cite{Scherrer2025}, freight flows, and modal interdependencies, as well as the more common scenario of partial degradation rather than complete failure. To address these gaps, this study applies a  network science framework to evaluate the structural and functional resilience of the U.S. intermodal freight transportation system, \rev{focusing} on rail and inland waterway components. An intermodal network is constructed using real infrastructure data and 2025 freight demand projections \cite{fhwa2022faf5,volpe2023}, integrating both topological structure and tonnage flow attributes.

\rev{Resilience is assessed through an iterative, scenario-based network simulation that evaluates system performance under progressive disruption. The approach simulates disruption scenarios in which nodes are degraded or removed according to predefined strategies.} Two primary disruption types are considered: (1) random failures, representing natural hazards \rev{(e.g., earthquakes)} or unforeseen breakdowns, and (2) targeted attacks on \rev{structurally or functionally critical nodes characterized by high traffic flow, high connectivity, or high centrality, and whose failure has a disproportionately large impact on the functionality of the overall network.} Unlike traditional models that assume complete node removal or failure, our framework also accounts for partial node failures\rev{,} offering a more realistic view of operational resilience.
The analysis in this study proceeds in three stages\rev{:}
\begin{enumerate}
    \item \textbf{Structural Characterization:} Identifying topological features such as degree distribution and spatial clusters of critical nodes to evaluate network redundancy.
    
    \item \textbf{Disruption Simulation:} Modeling the impact of both random and targeted node and edge removals on structural connectivity and freight-handling capacity.
    
    \item \textbf{Dynamic Degradation Modeling:} Examining how partial failures affect performance in terms of network efficiency.
\end{enumerate}

\noindent The objectives of this paper are as follows: (1) To investigate the structural characteristics of the U.S. intermodal freight transportation network; and (2) To evaluate both structural robustness and functional resilience under a range of disruption scenarios, including full and partial failures.
More broadly, \rev{our} work seeks to develop a scalable methodology for evaluating intermodal resilience and to generate actionable insights for infrastructure planners and policymakers. By identifying critical nodes  and modeling flow disruption, this study contributes to evidence-based strategies for enhancing the robustness, adaptability, and operational continuity of  freight networks.

\section{Literature Review}

The concept of resilience in transportation systems has evolved significantly beyond traditional reliability metrics, now encompassing the ability of infrastructure networks to absorb disruptions, maintain critical functions, and recover from adverse events \cite{wan2018resilience}. This expanded notion of resilience is particularly relevant in the context of intermodal freight systems, which are increasingly complex, spatially distributed, and interdependent across multiple transport modes \cite{chen2012resilience}.

Network science provides a robust analytical framework for studying such systems, allowing researchers to represent transportation infrastructures as graphs. {\rev{ Within this graph-theoretic paradigm, centrality measures quantify the relative importance of nodes or edges based on their position and role within the network, reflecting how traffic, connectivity, or information flows through the system.} Several studies have used topological measures such as degree, betweenness, and closeness centrality to identify vulnerable nodes and assess systemic robustness \cite{ crucitti2006centrality,holme2002attack, albert2000error}. These metrics reveal how connectivity patterns shape network vulnerability and redundancy, enabling the detection of high-impact elements.
In the resilience context, there is extensive literature exploring the effect of attacks on nodes \rev{with} high centrality \cite{rahimitouranposhti2025investigating, yadav2020resilience, bhatia2015network, chen2014assessing, cohen2001breakdown, callaway2000network}.

The foundational work by \cite{albert2000error} introduced the concept of error and attack tolerance in complex networks, showing that while scale-free networks are resilient to random failures, they are highly susceptible to targeted attacks on high-degree or high-betweenness nodes. This dual vulnerability framework has been widely applied to transportation studies and remains central to the analysis of network robustness under disruption. Extending this idea, \cite{holme2002attack} and \cite{bellingeri2014efficiency} emphasized the nonlinear effects of sequential node removals, demonstrating that the order and correlation of attacks significantly influence the speed and extent of fragmentation in both synthetic and real-world networks.

Disruption modeling in the literature is commonly categorized into random failure simulations and targeted attack scenarios \cite{aparicio2022assessing, yadav2020resilience}. While random disruptions emulate stochastic breakdowns or natural hazards, targeted removals prioritize elements based on static or adaptive centrality. The latter approach has been widely adopted due to its effectiveness in identifying critical nodes whose failure precipitates rapid connectivity loss. Recalculated centrality strategies further enhance this approach by dynamically updating node importance after each removal, offering more realistic assessments of evolving vulnerability.

Beyond traditional models assuming complete node removals, researchers have investigated the vulnerability of weighted networks, where link weights may represent flow volume, distance, or capacity. \cite{dall2006vulnerability} analyzed how the distribution of weights across links affects network breakdown, concluding that high-weight edges often serve as critical conduits for maintaining global efficiency. Their findings are especially pertinent for freight networks, where tonnage and route capacity vary substantially across corridors. Similarly, \cite{wang2011exploring} applied centrality-based vulnerability analysis to China’s air transport network, integrating both topological and flow-based metrics to identify functionally essential hubs. In an urban context, \cite{akbarzadeh2019role} demonstrated that travel demand, as a weight representing flow intensity, significantly influences the connectivity and resilience of street networks, with high-demand links being critical to maintaining system performance under disruptions.

Multimodal systems add another layer of complexity due to their inherent interdependencies. \cite{miller2012measuring} and \cite{mattsson2015vulnerability} highlighted that disruptions in one layer (e.g., rail) propagate through modal transfer points and amplify performance loss. This theoretical insight is supported by empirical evidence; for instance, \cite{aparicio2022assessing} assessed robustness in Lisbon’s multimodal system, demonstrating that modal integration improves reachability but also increases exposure to cascading disruptions. \rev{In a more recent study, Sun et al. \cite{sun2024network} proposed a topology-based framework to assess vulnerability to cascading failures in multimodal freight networks under limited data availability. Their approach relies on network topology and functional attributes, using weighted betweenness centrality as a proxy for infrastructure load and interdependencies, and is demonstrated through a regional case study of the Houston, Texas multimodal freight network.

While these studies have advanced understanding of multimodal vulnerability and robustness, important gaps remain for flow-explicit, national-scale resilience analysis of real-world intermodal freight networks, particularly in the U.S. Existing studies primarily emphasize vulnerability and cascading failures at regional scales using topological proxies, rather than directly modeling freight demand, and functional performance degradation under disruption.

To address these limitations, this study evaluates both the structural and functional resilience of the U.S. intermodal freight transportation system by explicitly integrating network topology with tonnage-based flow attributes derived from 2025 freight demand projections. Focusing on rail and inland waterway networks at the national scale, the proposed framework enables direct assessment of disruption impacts on freight flows, and system-level performance under partial and full degradation scenarios. The contribution of this study lies in systematically combining flow-weighted metrics and progressive degradation modeling within a unified intermodal framework that explicitly captures tonnage-based demand, modal interdependencies, and partial degradation effects, thereby providing actionable insights for resilience planning in U.S. rail and inland waterway freight systems.}

\section{Research Methodology}
\subsection{Data Collection and Processing}

The intermodal freight network is modeled as an \rev{origin-destination (OD)} system comprising U.S. rail and water infrastructure. Railway nodes are selected from Norfolk Southern and Union Pacific intermodal terminals, totaling 55 locations identified by their \rev{geographic} coordinates, while waterway nodes include 47 major inland and coastal ports, chosen based on waterborne tonnage data from the U.S. Army Corps of Engineers and the Freight Analysis Framework (FAF) \cite{fhwa2022faf5}. \rev{All edges, including rail and water connections, are constructed using the Freight and Fuel Transportation Optimization Tool (FTOT) \cite{volpe2023}, which generates feasible transportation links based on existing infrastructure topology, geographic connectivity, and mode compatibility. Rail edges follow established rail corridors associated with the Norfolk Southern and Union Pacific networks, while waterway edges are defined along navigable inland rivers and coastal shipping routes. This network captures the backbone infrastructure that carries the majority of U.S. intermodal freight flow, making it appropriate for analyzing system-level resilience and identifying critical chokepoints.} Freight demand is extracted from FAF using 113 U.S. cities and states as \rev{OD} pairs and includes projected tonnages for the year 2025. \rev{Since} the FAF demand projections primarily reflect truck-based flows, we apply a proportional scaling approach to estimate equivalent rail and water freight volumes. Based on \rev{the} 2017 national modal share statistics, road accounts for approximately 82\% of total freight tonnage, while rail and water represent 12\% and 6\%, respectively \cite{BTS2017FreightMode}. These shares, derived from total freight volumes of 11.5 million (truck), 1.74 million (rail), and 0.77 million (water) \rev{measured in} thousand tons, are used to scale each OD flow using adjustment factors of 0.124 for rail and 0.055 for water.

\rev{ We note that the use of uniform national modal share factors does not aim to reproduce corridor- or commodity-specific modal splits, which are known to vary substantially across geography and freight type. Instead, this proportional scaling serves as a controlled approximation to generate a demand-weighted network for large-scale resilience analysis, where the primary interest lies in relative performance degradation, and robustness rankings under disruption rather than precise mode choice estimation. By preserving the spatial distribution and relative magnitudes of FAF demand, this approach ensures that high-volume corridors and hubs remain dominant in the analysis, enabling consistent comparison of network behavior across disruption scenarios. Similar  simplifying assumptions are commonly adopted in system-level resilience and vulnerability studies when detailed mode-specific OD data are unavailable \cite{mattsson2015vulnerability, sun2024network}.} Other modes\rev{,} such as pipeline, air, and multiple-mode shipments\rev{,} were excluded from the normalization to ensure comparability across surface-based freight modes and to reflect realistic modal substitution patterns within the intermodal network scope. This approach preserves the spatial structure of demand while providing a practical approximation of non-road flows in the absence of detailed mode-specific OD data, with the assumption of uniform modal substitution across pairs. For OD pairs with missing demand data, a geospatial fallback mechanism was implemented to identify the nearest intermodal hub with known demand using Euclidean distance. If no direct match was available, the search was extended to the second-nearest or a cross-state alternative within the same region. This approach enabled partial recovery of otherwise missing demand values and improved the completeness of the OD matrix for simulation.

\subsection{Method}

This study investigates the structural characteristics and resilience of the U.S. intermodal freight network through a combination of graph-theoretic and flow-based approaches. The analysis is divided into three core components: (1) structural characterization and network topology analysis, (2) simulation of node and edge failure scenarios to assess structural robustness, and (3) evaluation of functional resilience using flow-weighted centrality metrics and partial degradation.

\subsubsection{Structural Characterization and Network Topology}

We begin by analyzing the network's topology using degree-based metrics. The degree of a node \( i \), denoted \( k_i \), is defined as the number of links connected to it. This measure helps characterize the underlying connectivity pattern and detect hub structures. A mode-aware version of this analysis further disaggregates degree by transport type (rail and waterway), highlighting modal heterogeneity. The cumulative degree distribution \( P(K \geq k) \) is then calculated to represent the probability that a randomly selected node has a degree greater than or equal to \( k \). This provides a smoothed perspective of node connectivity, particularly useful in identifying highly connected hubs. 

\subsubsection{Structural Robustness Under Disruption Scenarios}

To evaluate network robustness, we simulate node and edge removals under five strategies: random, initial degree-based (ID), initial betweenness-based (IB), and their adaptive counterparts, recalculated degree (RD) and recalculated betweenness (RB). \rev{For the random strategy, a random node order is generated (using a fixed seed for reproducibility), whereas the targeted strategies use deterministic centrality rankings.} 

We represent the intermodal freight network as a graph \( G = (V, E) \), where \( V \) is the set of nodes (e.g., rail stations, ports, terminals), and \( E \) is the set of edges representing direct transport links between nodes. In the ID strategy, nodes are removed in descending order of their initial degree centrality. \rev{For each node \( i \in V \), the degree centrality \( k_i \) \cite{freeman1978centrality} is defined in Eq. \ref{degree} as:}

\begin{equation} \label{degree}
k_i = \sum_{j \in V} a_{ij}
\end{equation}

\noindent where \( a_{ij} = 1 \) if there is an edge between nodes \( i \) and \( j \), and \( 0 \) otherwise.

\noindent For IB removal, nodes are removed based on their betweenness centrality \cite{freeman1978centrality}, \rev{defined in Eq. \ref{betweenness} as:}

\begin{equation} \label{betweenness}
c_B(i) = \sum_{s \neq t \neq i \in V} \frac{\sigma(s, t \mid i)}{\sigma(s, t)}
\end{equation}

\noindent where \( \sigma(s, t) \) denotes the number of shortest paths between nodes \( s \) and \( t \), and \( \sigma(s, t \mid i) \) denotes the number of those paths that pass through node \( i \). The RD and RB strategies recalculate the respective centrality measures after each removal, better representing dynamic attack scenarios \cite{holme2002attack}. These strategies are extended to edge removals to reflect infrastructure vulnerabilities such as bridge or tunnel failures. 

\subsubsection{Functional Resilience Using Flow-Based Centralities}

To assess the functional resilience of the intermodal freight network, we incorporate demand flow data into the centrality measures. Two key flow-weighted centrality metrics are used: Weighted Degree Centrality (WDC) and Weighted Betweenness Centrality (WBC) \cite{jing2025critical}. 

WDC quantifies the total volume of inbound and outbound tonnage at a node, capturing its local importance as a freight hub. 
For a given node \( i \in V \), let 
$N_{\mathrm{in}}(i) = \{\, j \in V \mid (j,i) \in E \,\}$
denote the set of nodes with edges directed into \(i\), and 
$N_{\mathrm{out}}(i) = \{\, j \in V \mid (i,j) \in E \,\}$ denote the set of nodes with edges directed out from \( i \). WDC is defined in Eq. \ref{eq:WD} as:

\begin{equation}
\label{eq:WD}
\mathrm{WD}(i)
= \sum_{j \in N_{\mathrm{in}}(i)} \theta_{ji}
+ \sum_{j \in N_{\mathrm{out}}(i)} \theta_{ij}
\end{equation}

\noindent where \( \theta_{ji} \) represents the inbound freight tonnage from node \( j \) to \( i \), and \( \theta_{ij} \) denotes the outbound tonnage from \( i \) to node \( j \). This metric reflects the volume of freight a node processes, regardless of its network position.

To capture the role of a node in facilitating freight movement across the network, we use WBC. This metric extends the traditional betweenness centrality by incorporating demand weights and is expressed \rev{ in Eq. \ref{WB} as:}

\begin{equation} \label{WB}
WB(i) = \frac{1}{\sum_{s,t \in V} W_{st}} \sum_{s,t \in V} \frac{\sigma_{st}(i)}{\sigma_{st}} \cdot W_{st}
\end{equation}

\noindent where \( \sigma_{st} \) is the number of shortest paths between nodes \( s \) and \( t \), \( \sigma_{st}(i) \) is the number of those paths that pass through node \( i \), and \( W_{st} \) denotes the freight demand (in tons) between \( s \) and \( t \). The entire expression is normalized by the total flow \( \sum_{s,t \in V} W_{st} \), ensuring comparability across different network sizes and traffic intensities. Under both WDC and WBC removal strategies, nodes are eliminated in descending order of their centrality values.

In addition to complete node removals, we also consider partial failures, which reflect more realistic forms of infrastructure degradation. Instead of assuming binary functionality (fully operational or entirely disabled), we simulate reduced capacity at critical nodes. This allows us to model disruptions such as labor slowdowns, congestion, or \rev{partial shutdown scenarios} in which the infrastructure continues to operate, but at a diminished level of service. \rev{Under partial failure scenarios, affected nodes retain their network connections, but the effective freight throughput on their incident links is proportionally reduced, enabling the analysis to capture graded performance loss rather than abrupt topological disconnection.}

\rev{Partial degradation is simulated by scaling the freight flow (tonnage) associated with all edges incident to a degraded node. For a node operating at a reduced functionality level $\alpha \in \{0.7, 0.4, 0.0\}$, the tonnage carried on each incoming and outgoing edge connected to that node is multiplied by $\alpha$. The same scaling is applied to the corresponding origin-destination demand contributions that traverse these edges. This approach captures reduced processing or transfer capacity at the node without removing its physical connections. Degradation is applied sequentially to selected nodes according to the specified disruption strategy. Both network topology and edge travel costs remain unchanged, and shortest paths are recomputed on the original distance-weighted network at each simulation step.}

\subsubsection{Resilience Metrics}

Network resilience is evaluated using two primary metrics: one that captures structural resilience and another that evaluates functional resilience under complete and partial degradation. For structural resilience, we monitor the  Giant Connected Component (GCC) as nodes or edges are progressively removed. The GCC is defined as the largest subset of nodes  in which a path connects every pair of nodes, that is, the largest maximally connected component \cite{bollobas2011random}. It captures the core structure of the network that remains functionally cohesive under disruption. Let \( f \) denote the fraction of removed elements, and let \( GCC_f \) represent the size of the largest connected component after the removal of fraction \( f \). The resilience curve is computed \rev{ in Eq. \ref{GCC} as:}

\begin{equation}\label{GCC}
S(f) = \frac{|GCC_f|}{|V|}
\end{equation}
\noindent where \( |V| \) is the total number of nodes in the original network. This normalized measure allows us to track how connectivity deteriorates as failures accumulate. 

While structural resilience captures the preservation of connectivity under disruptions, functional resilience reflects the system’s ability to maintain performance. To measure functional resilience, we employ the flow-weighted Network Efficiency (NE) metric \cite{latora2002boston}\rev{, which evaluates how effectively freight demand can be transported across the network given its current topology and link distances. This measure quantifies the system's demand-weighted functional efficiency, accounting for both the volume of demand and the cost of movement between origin-destination pairs.} It is formally defined \rev{ in Eq. \ref{eq:NE} as:}

\begin{equation}
\label{eq:NE}
\mathrm{NE}_{\mathrm{weighted}}
= \frac{1}{|V| \, (|V| - 1)}
\sum_{i \in V}
\sum_{j \in V \setminus \{i\}}
\frac{w_{ij}}{d_{ij}}
\end{equation}

\noindent where \( w_{ij} \) denotes the freight demand (in tons) between nodes \( i \) and \( j \), and \( d_{ij} \) is the shortest path distance between them \rev{computed on the weighted network. In this study, the edge weight used to compute $d_{ij}$ corresponds to the physical link distance (miles) assigned to each rail or waterway connection.} The expression is normalized by the total number of node pairs \( |V|(|V| - 1) \), ensuring comparability across networks of varying sizes \rev{ and enabling direct assessment of how disruptions degrade the network's ability to efficiently move freight. Higher NE values indicate that large demand flows are being served via short, efficient routes, while lower values reflect either demand being transported over longer paths or the complete loss of connectivity for some origin-destination pairs, which then make no contribution to the summation. Under this formulation, NE represents a demand-weighted measure of mobility efficiency, reflecting how easily freight can be moved across the network as disruptions alter connectivity or increase effective travel distances. Reductions in NE therefore indicate loss of functional performance due to longer routing paths or disconnected origin–destination pairs, rather than reductions in physical capacity or throughput.}

\section{Results}

\subsection{Structural Characteristics of the Network}

The \rev{GCC} of the undirected intermodal freight network analyzed in this study \textbf{(Figure~\ref{fig:network})} comprises 78 nodes and 179 edges. \rev{The full constructed intermodal network contains 102 nodes in total (55 rail terminals and 47 ports), of which the GCC represents the largest connected component. The remaining 24 nodes exist as isolated facilities or small disconnected components and are therefore excluded from the resilience analysis.} Within this component, the rail subnetwork contains 55 nodes and 125 edges, while the waterway subnetwork consists of 47 nodes and 64 edges. \textbf{Figure~\ref{fig:network}} also distinguishes between single-mode and dual-mode nodes and edges. A total of 14 nodes are connected by dual-mode edges\rev{. These nodes represent} the key intermodal transfer points where freight can switch between rail and water modes. \rev{These 14 dual-mode nodes are not an additional set, but rather a subset of the 78 GCC nodes that serve as intermodal transfer points with connections to both rail and waterway edges.}

\begin{figure}[!ht]
  \centering
  \includegraphics[width=0.9\textwidth]{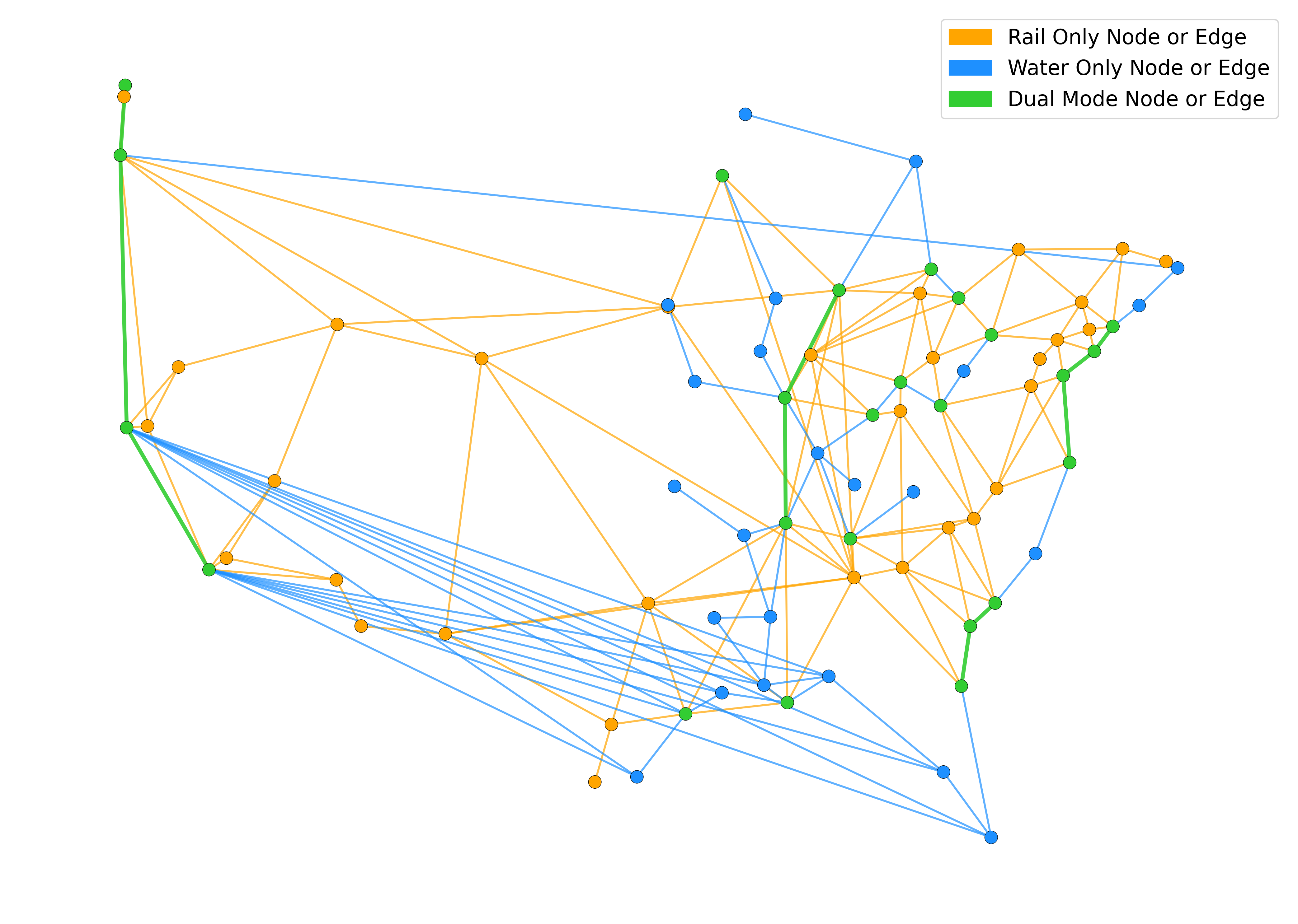}
  \caption{\textbf{Undirected U.S. intermodal freight network illustrating spatial connectivity by mode. 
  Orange denotes rail-only infrastructure, blue indicates water-only, and green represents dual-mode nodes and edges that support both rail and water transport.}}
  \label{fig:network}
\end{figure}

To evaluate the overall connectivity of the network, we analyze the degree distribution of the GCC. \textbf{Figure~\ref{fig:degree}} displays the distribution of node degrees, representing the number of direct connections each node has. \textbf{Figure~\ref{fig:cumu degree}} shows the cumulative degree distribution (CDD), which provides a more integrated view of the network’s connectivity and helps identify the presence of hub nodes. The average degree of the combined intermodal GCC is 4.59. This value is comparable to other large-scale infrastructure networks such as the World Wide Web, which has an average degree of approximately 4.60 \cite{posfai2016network}. When examining each transport mode separately, the rail subnetwork shows a higher average degree of 4.55, while the waterway subnetwork has a notably lower average degree of 2.72. This lower connectivity among water-only nodes is expected, given the geographic constraints of water-based infrastructure. Waterway routes are often confined to coastlines and river systems, and many water-only nodes serve as feeder ports or local terminals rather than as regional hubs.




\begin{figure}[!ht]
\centering

\begin{subfigure}[b]{0.45\textwidth}
\centering
\begin{tikzpicture}
\begin{axis}[
    width=\textwidth,
    height=7cm,
    xlabel={Node Degree},
    ylabel={Frequency},
    xmin=0.5, xmax=14.5,
    ymin=0, ymax=16,
    xtick={2,4,6,8,10,12,14},
    ytick={0,2,4,6,8,10,12,14,16},
    grid=both,
    grid style={dashed, gray!30},
    major grid style={dashed, gray!50},
    xlabel style={font=\normalsize},
    ylabel style={font=\normalsize},
    tick label style={font=\small},
    ybar,
    bar width=11.5 pt, 
    enlarge x limits=0.02,
]
\addplot[
    fill=cyan!35,
    draw=black,
    line width=0.3pt,
] coordinates {
    (1, 7)
    (2, 11)
    (3, 8)
    (4, 10)
    (5, 15)
    (6, 13)
    (7, 4)
    (8, 4)
    (9, 1)
    (10, 1)
    (11, 1)
    (12, 1)
    (13, 2)
};
\end{axis}
\end{tikzpicture}
\caption{Node Degree Distribution}
\label{fig:degree}
\end{subfigure}
\hfill
\begin{subfigure}[b]{0.45\textwidth}
\centering
\begin{tikzpicture}
\begin{axis}[
    width=\textwidth,
    height=7cm,
    xlabel={Degree (k)},
    ylabel={No. of Nodes with Degree $\geq$ k},
    xmin=0.5, xmax=14,
    ymin=0, ymax=80,
    xtick={2,4,6,8,10,12,14},
    ytick={0,10,20,30,40,50,60,70,80},
    grid=both,
    grid style={dashed, gray!30},
    major grid style={dashed, gray!50},
    xlabel style={font=\normalsize},
    ylabel style={font=\normalsize},
    tick label style={font=\small},
]
\addplot[
    color=blue,
    mark=*,
    mark size=2pt,
    thick,
] coordinates {
    (1, 78)
    (2, 71)
    (3, 60)
    (4, 52)
    (5, 42)
    (6, 27)
    (7, 14)
    (8, 10)
    (9, 6)
    (10, 5)
    (11, 4)
    (12, 3)
    (13,2)
};
\end{axis}
\end{tikzpicture}
\caption{Cumulative Degree Distribution}
\label{fig:cumu degree}
\end{subfigure}

  \caption{\textbf{Degree distribution of the intermodel modal transportation network comprising rail and water modes: (a) Degree distribution of the GCC; (b) Cumulative degree distribution of the GCC.
  }}
  \label{fig:degree all}
\end{figure}
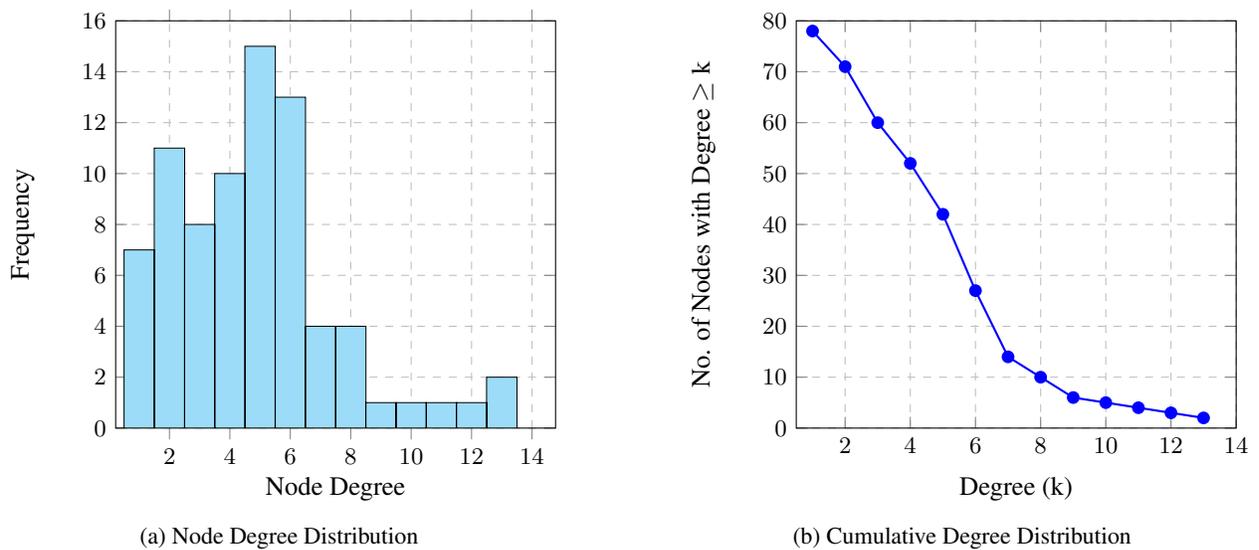

The cumulative degree distribution \textbf{(Figure~\ref{fig:cumu degree})} further highlights structural characteristics of the network. While the network is moderately connected \rev{overall, node degree values decline sharply beyond a degree of five. Only about 35 nodes, approximately 45\% of the total 78 nodes in the \rev{GCC,} have a degree greater than five,} forming a small core that likely supports intermodal connectivity. The fact that most nodes have four or more connections suggests a degree of robustness to random failures, where isolated disruptions are unlikely to fragment the network. However, the distribution also reveals potential vulnerability to targeted attacks. Only four nodes have degrees of ten or higher, and just two exceed twelve, indicating that a small set of super-hubs plays a disproportionately critical role. Removing one of these could significantly disrupt overall connectivity. Redundancy appears moderate, with 24 nodes having six or more connections, which could allow rerouting and mitigating some risks.
The spatial distribution of these high-degree nodes is particularly important in evaluating the network’s robustness. A concentrated cluster of critical nodes could amplify the impact of regional disruptions, whereas dispersion increases the likelihood of maintaining alternative pathways. As shown in \textbf{Figure~\ref{fig:degree_greater_than_6}}, these nodes are not spatially clustered but rather spread across different geographic regions. This dispersion may enhance the network’s ability to absorb localized disruptions and maintain operational continuity by enabling flexible rerouting.

\begin{figure}[!htbp]
\centering
 \includegraphics[width=0.85\textwidth]{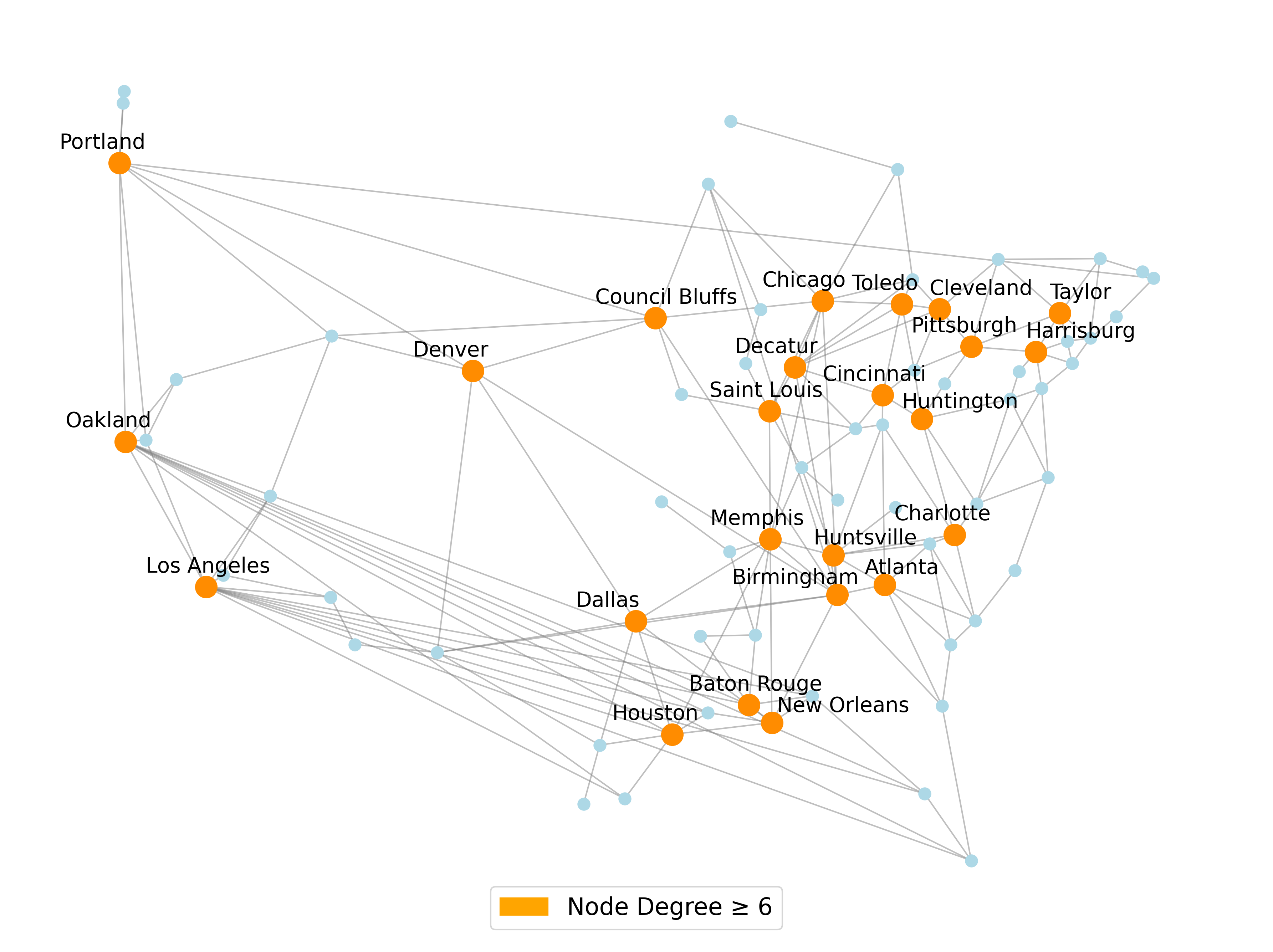}
    \caption{\textbf{Geographic distribution of high-degree nodes illustrating spatial redundancy}}
    \label{fig:degree_greater_than_6}
\end{figure}

\subsection{Structural Resilience of the Network}

To assess the structural resilience of the intermodal freight network, five disruption strategies were applied under both node and edge removal scenarios. The impacts were evaluated by tracking changes in GCC size and the number of isolated components. Among all approaches, \rev{RB} proves to be the most effective at disrupting network connectivity. As shown in \textbf{Figure~\ref{fig:remove nodes}}, removing just five nodes or edges under the RB strategy reduces the GCC size to nearly half its original value. This strategy continues to identify and remove structurally critical elements throughout the process, making it the most aggressive in fragmenting the network.

\begin{figure}[!ht]
  \centering
  \begin{subfigure}[b]{0.48\textwidth}
    \centering
    \includegraphics[width=\textwidth]{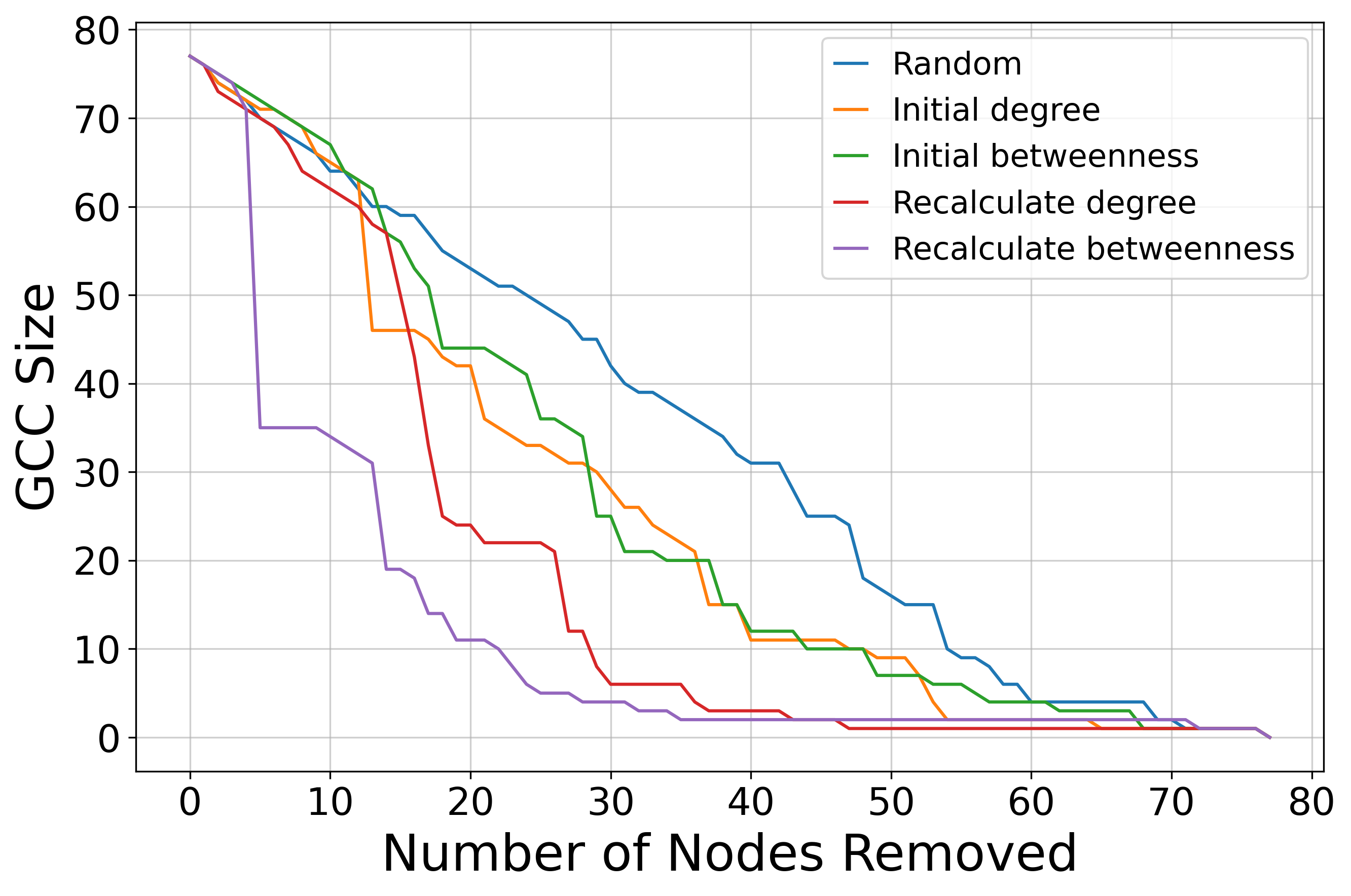}
    \caption{}
    \label{fig:remove nodes}
  \end{subfigure}
  \hfill
  \begin{subfigure}[b]{0.48\textwidth}
    \centering
    \includegraphics[width=\textwidth]{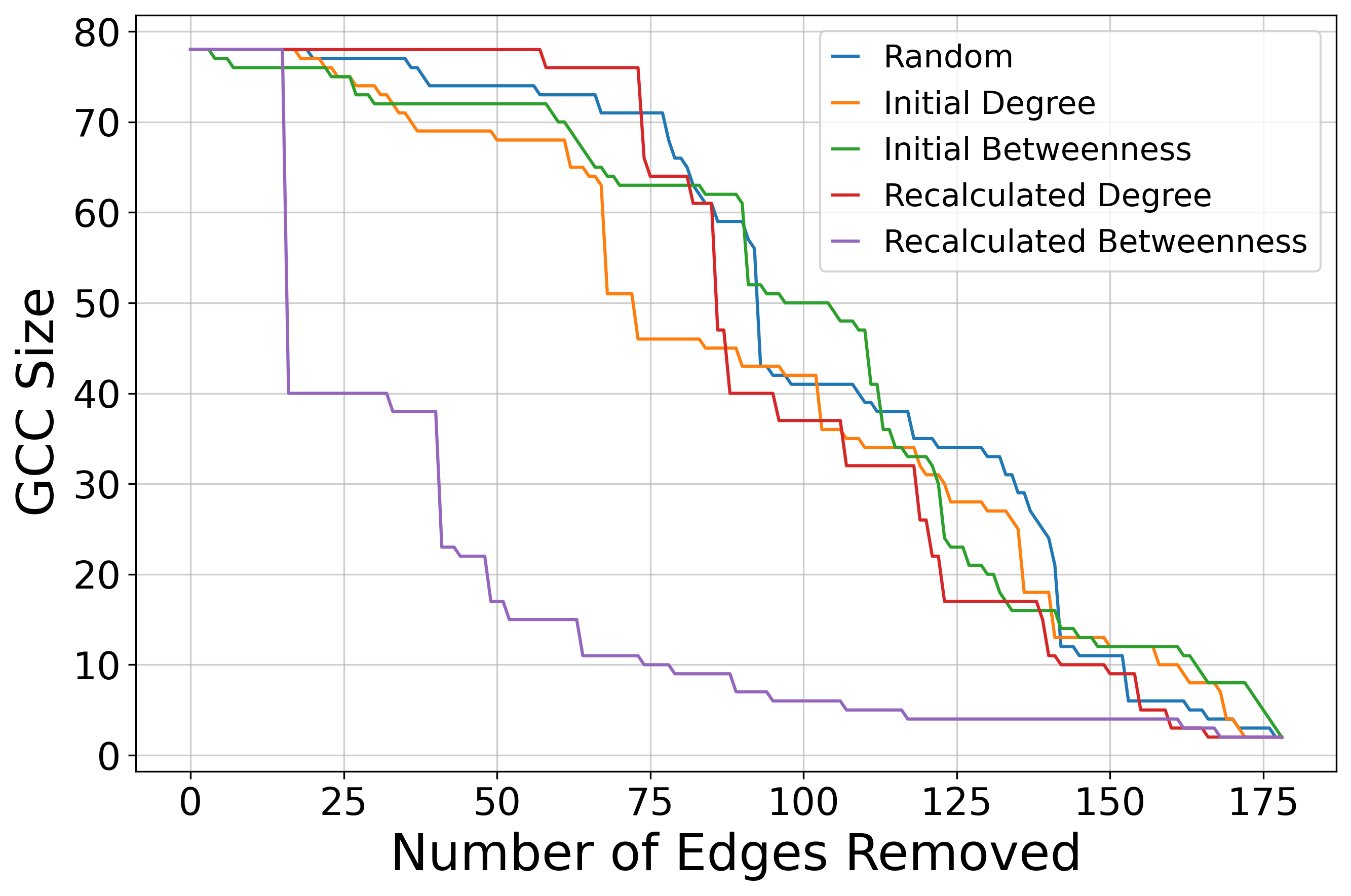}
    \caption{}
    \label{fig:remove edges}
  \end{subfigure}
  
  \caption{\textbf{Evolution of the Giant Connected Component (GCC) size under different disruption scenarios: 
  (a) node removals and (b) edge removals.}}
  \label{fig:gcc removal}
\end{figure}

For node removal \textbf{(Figure~\ref{fig:remove nodes})}, the \rev{ID} strategy causes greater early fragmentation, with noticeable impact around 15 node removals. However, \rev{RD} becomes more disruptive thereafter, as it continuously adapts to the evolving network and targets newly critical nodes, leading to greater fragmentation over time. For edge removal \textbf{(Figure~\ref{fig:remove edges})}, the two strategies diverge in behavior. \rev{RD} maintains a nearly intact GCC until about 60 edge removals, after which it triggers a sharp decline. In contrast, \rev{ID} begins fragmenting the network earlier, around 20 edge removals, but follows a slower and more gradual decline. Although \rev{ID} initially reduces the GCC more effectively, it is eventually surpassed by \rev{RD}. This pattern suggests that static centrality measures are more effective in the early stages, while adaptive strategies become increasingly disruptive as the network structure changes. \rev{Static rankings primarily identify the most important hubs in the intact network, whereas adaptive strategies continuously update node importance to capture newly emergent bottlenecks as connectivity and flows are redistributed following successive failures.}


\begin{figure}[!ht]
  \centering
  \begin{subfigure}[b]{0.48\textwidth}
    \centering
    \includegraphics[width=\textwidth]{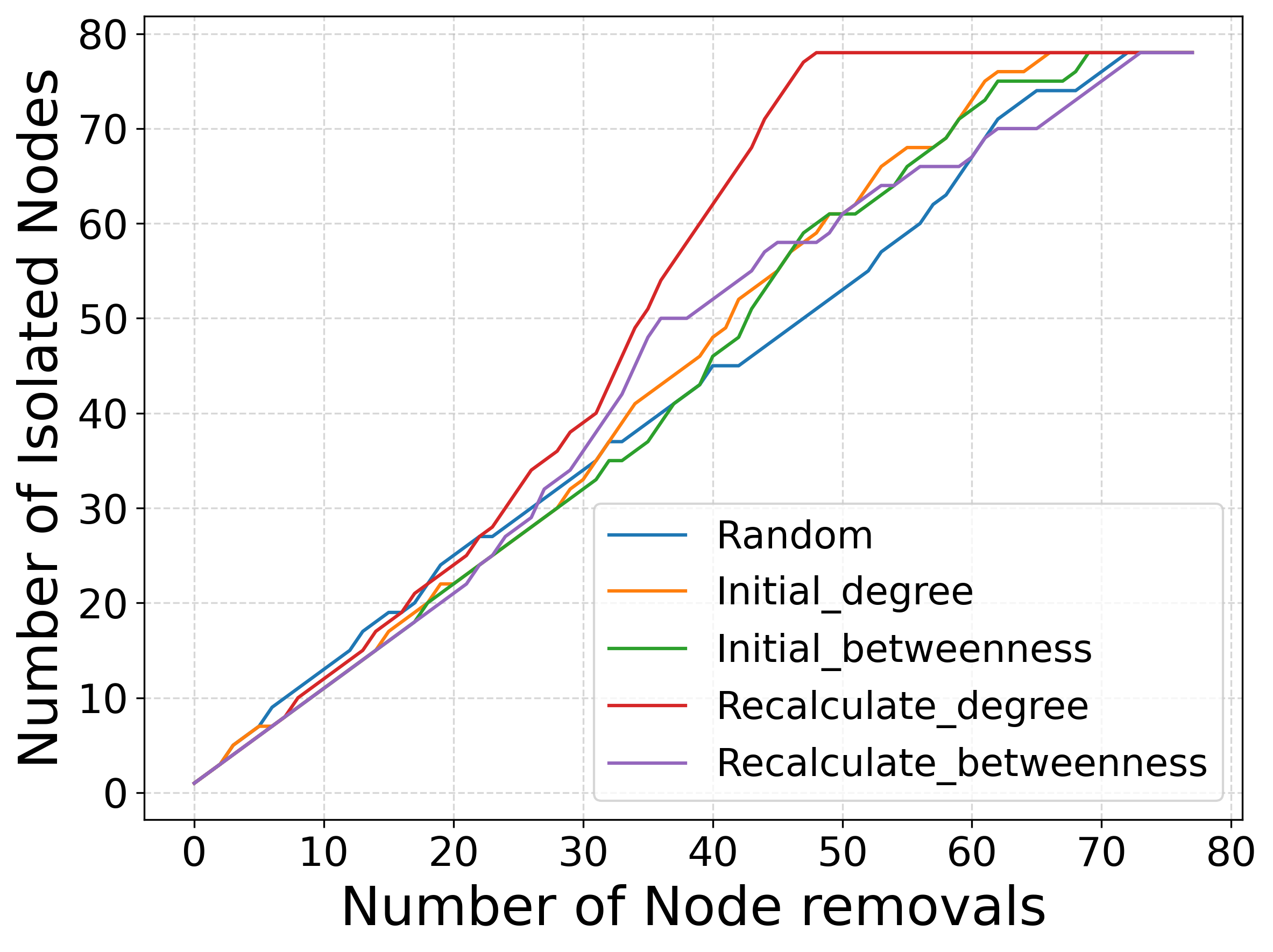}
    \caption{}
    \label{fig:isolated nodes}
  \end{subfigure}
  \hfill
  \begin{subfigure}[b]{0.48\textwidth}
    \centering
    \includegraphics[width=\textwidth]{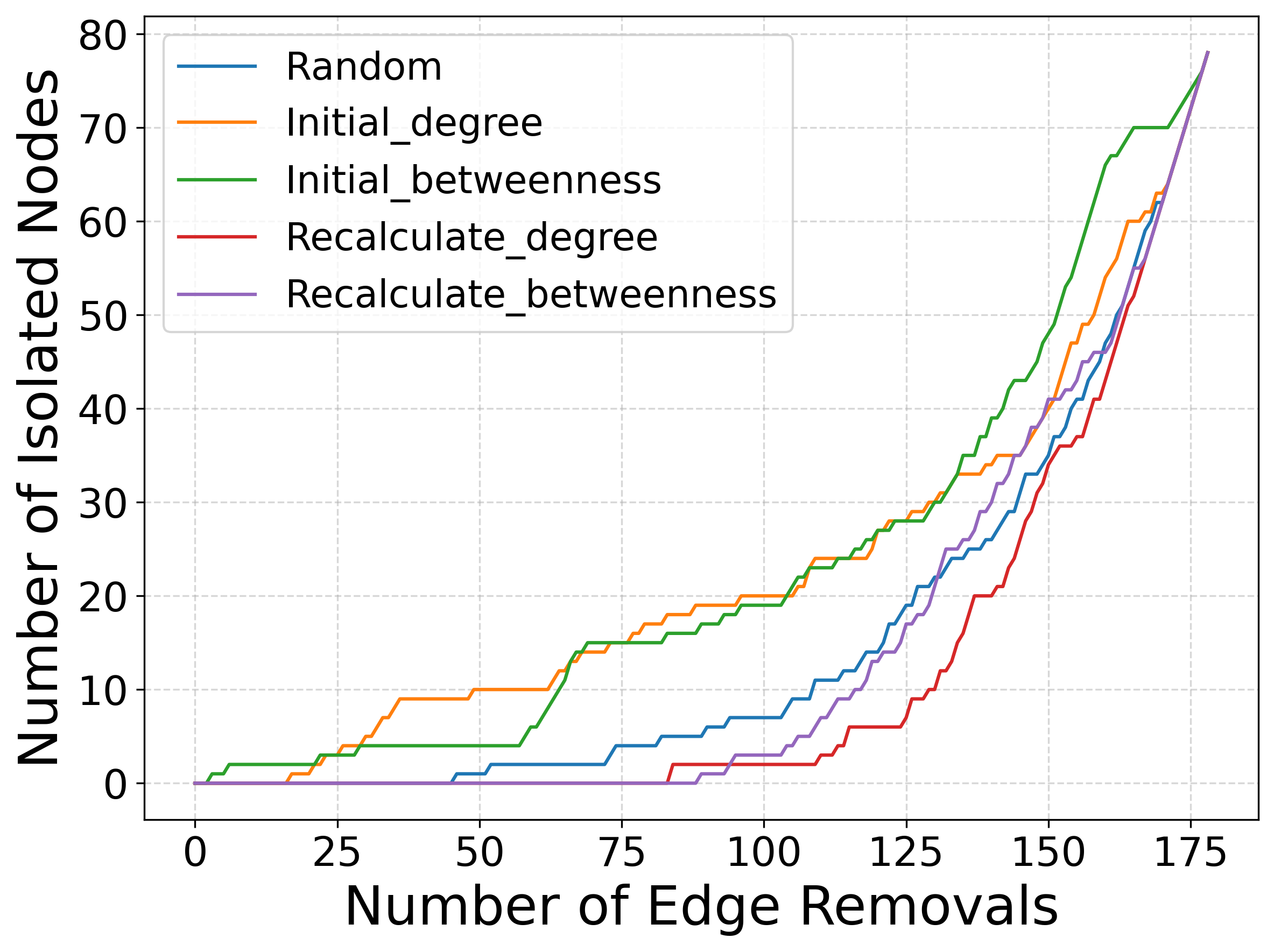}
    \caption{}
    \label{fig:isolated edges}
  \end{subfigure}
  
  \caption{\textbf{Evolution of isolated components under different disruption scenarios: 
  (a) node removals and (b) edge removals.}}
  \label{fig:isolated components}
\end{figure}

The analysis of isolated components offers additional insight into network fragmentation. \rev{An isolated component forms when a node loses all its edges.} \textbf{Figures~\ref{fig:isolated nodes} and~\ref{fig:isolated edges}} illustrate the evolution of isolated components for node and edge removal, respectively. All strategies, except \rev{RD}, cause a steady increase in isolated nodes, reflecting ongoing fragmentation. \rev{RD} stands out for its sharp tipping point as it requires approximately 48 node removals \textbf{(Figure~\ref{fig:isolated nodes})} to fully fragment the network into isolated components. This makes it the most efficient strategy for total network disconnection via node removal. In contrast, the other four strategies increase isolation \rev{more gradually}, suggesting a different pattern of failure with slower full fragmentation. For edge removal, the \rev{RD and RB} strategies exhibit minimal impact until approximately 80 edges are removed \textbf{(Figure~\ref{fig:isolated edges})}. This delayed response suggests that these dynamically updated strategies initially fail to target structurally critical edges, likely because recalculated metrics at early stages prioritize peripheral or redundant links. However, beyond this threshold, both strategies trigger a rapid fragmentation of the network, indicating a delayed but abrupt structural breakdown once key connectors are eventually exposed. In contrast, the \rev{ID and IB} strategies are far more effective from the outset. By targeting edges based on their original centrality values, they immediately disrupt critical shortest-path corridors and initiate progressive network disintegration. Notably, these initial strategies continue to outperform both recalculated variants throughout the edge removal process. While \rev{ID} is more impactful in the early stages, \rev{IB} overtakes it after approximately 130 edge removals and remains the most effective strategy thereafter.

In prior robustness analyses \textbf{(Figures~\ref{fig:gcc removal} and~\ref{fig:isolated components})}, edge weights, defined as physical distances between nodes, were incorporated into centrality calculations to reflect real-world freight dynamics, where longer links imply higher transport costs or travel time. To evaluate whether incorporating distance-based weights improves the identification of critical nodes, we compare weighted and unweighted versions of \rev{degree- and betweenness-based node removal} in \textbf{Figures~\ref{fig:degree-weighted-vs-unweighted} and~\ref{fig:betweenness-weighted-vs-unweighted}}. For degree centrality, the unweighted measure simply counts the number of directly connected edges, while the weighted variant sums the total distance of those links, effectively capturing the spatial strength of a node’s connectivity. For betweenness, the difference is even more pronounced: without weights, shortest paths are based on hop count alone, whereas the weighted version identifies paths with the lowest cumulative distance, offering a more realistic view of transit-critical nodes. The results show that weighted strategies more rapidly reduce the size of the GCC compared to their unweighted counterparts, indicating that incorporating distance yields more accurate assessments of network vulnerability and better reflects the real-world significance of nodes in freight connectivity.

\begin{figure}[!ht]
  \centering
  \begin{subfigure}[b]{0.48\textwidth}
    \centering
    \includegraphics[width=\textwidth]{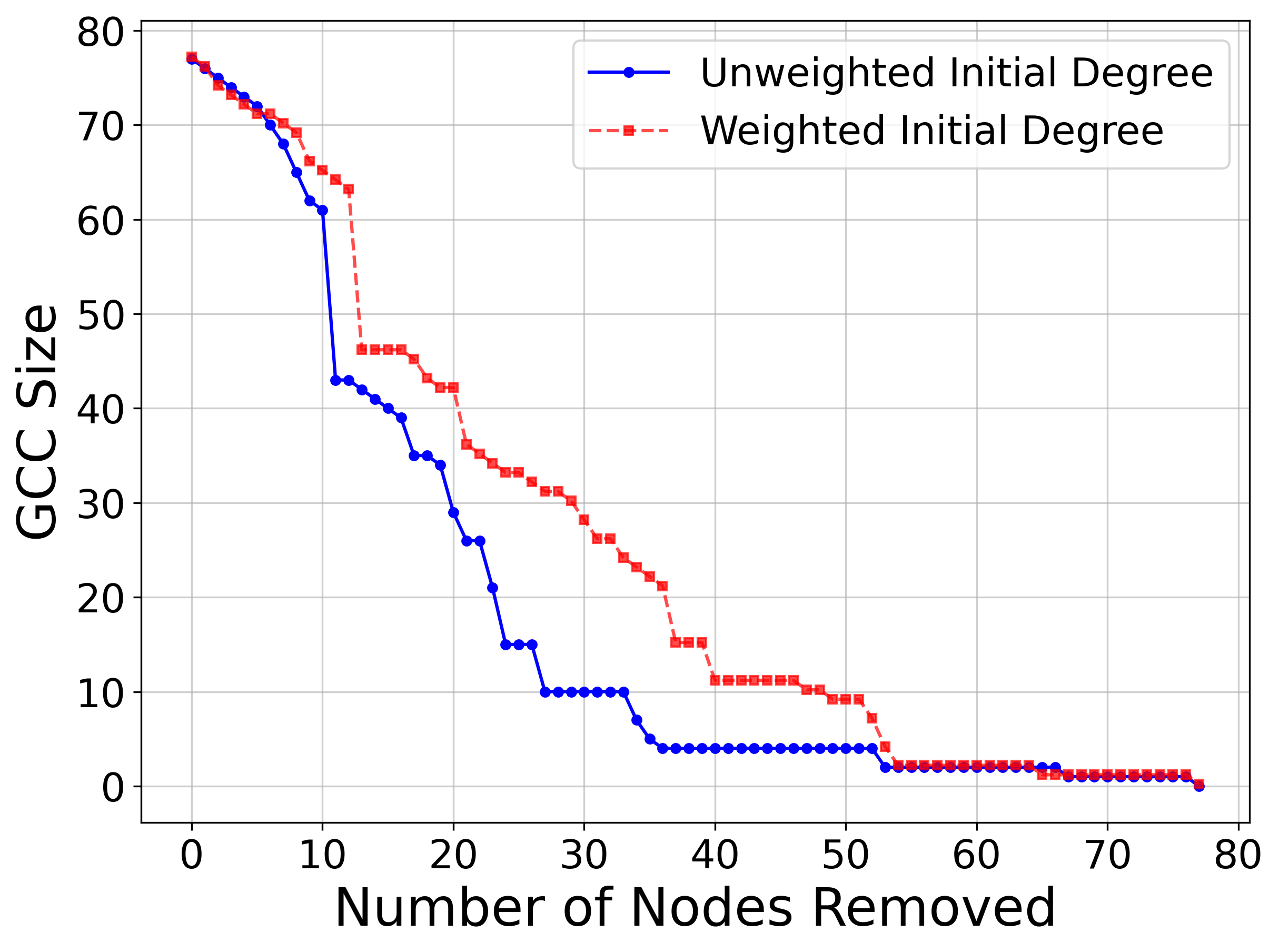}
    \caption{}
    \label{fig:degree-weighted-vs-unweighted}
  \end{subfigure}
  \hfill
  \begin{subfigure}[b]{0.48\textwidth}
    \centering
    \includegraphics[width=\textwidth]{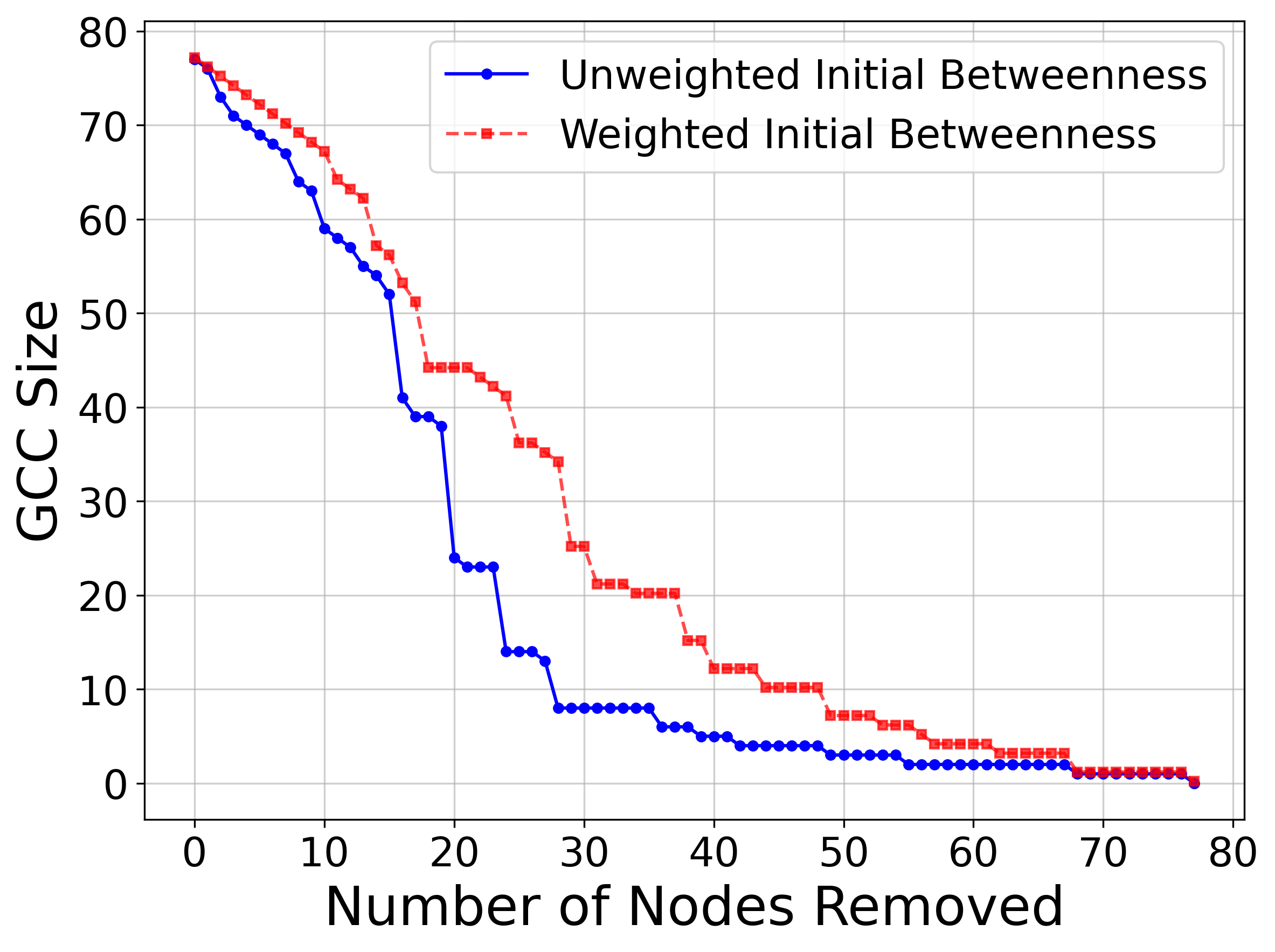}
    \caption{}
    \label{fig:betweenness-weighted-vs-unweighted}
  \end{subfigure}
  
  \caption{
  \textbf{Impact of weighted vs. unweighted strategies on GCC size under targeted node removal: 
  (a) ID-based and (b) IB-based. 
  }}
  \label{fig:weighted-vs-unweighted}
\end{figure}

Another key comparative insight is that node removals are significantly more damaging than edge removals. Achieving a similar reduction in GCC size or increase in isolated components requires far more edge removals. \rev{This pattern is evident in real-world freight disruptions: prolonged closures of major ports due to labor strikes, equipment failures, or operational shutdowns (node disruptions) typically generate more severe and widespread supply chain impacts than temporary bridge closures or track maintenance shutdowns (edge disruptions), even when multiple links are affected simultaneously \cite{mattsson2015vulnerability}.} This is expected, as removing a node also eliminates all its incident edges, whereas edge removal affects only part of a node's connectivity. Additionally, node removals are more likely to sever structural backbones, while edge removals often leave alternative routes intact due to network redundancy. Even when critical edges are removed, many destinations remain reachable through longer paths, making edge-based disruptions less immediately impactful unless they occur as part of broader regional or coordinated failures.


\subsection{Functional Resilience of the Network}
To assess the system’s functional resilience, demand-weighted flow data is integrated into two flow-sensitive centrality measures: \rev{WDC and WBC}. These metrics capture a node’s importance in facilitating flow across the network.
\textbf{Figure~\ref{fig:failure}} presents the degradation (failure phase) of normalized weighted \rev{NE} under different node removal strategies. During the failure phase, WDC-based attacks cause the most rapid decline in NE, implying that nodes with high cumulative freight volumes play an outsized role in maintaining efficient system-wide flow. Their removal directly eliminates large portions of traffic, regardless of whether these nodes serve as connectors between other regions. In contrast, WBC-based failures, while still targeted, result in a more gradual degradation curve. Interestingly, the WDC and random failure curves intersect multiple times, indicating alternating dominance in their impact. \rev{The intersections occur because random failures can occasionally remove high-throughput or flow-critical nodes earlier than their position in the WD ranking, while differences in freight volumes among mid-ranked WD nodes are relatively small, causing stage-dependent reversals in relative impact that depend on which OD flows are disrupted.} For example, WDC-based attacks are more damaging at 10 and 30 node removals, while random failures result in greater NE loss at 20 and 40 \rev{removed nodes.} Beyond roughly 42 removals, WDC becomes consistently more destructive. These fluctuations suggest that although random failures occasionally strike high-impact nodes by chance, only WDC-based targeting consistently accelerates functional degradation by prioritizing nodes with the highest cumulative freight load.


\begin{figure}[H]
\centering
  \includegraphics[width=0.7\textwidth]{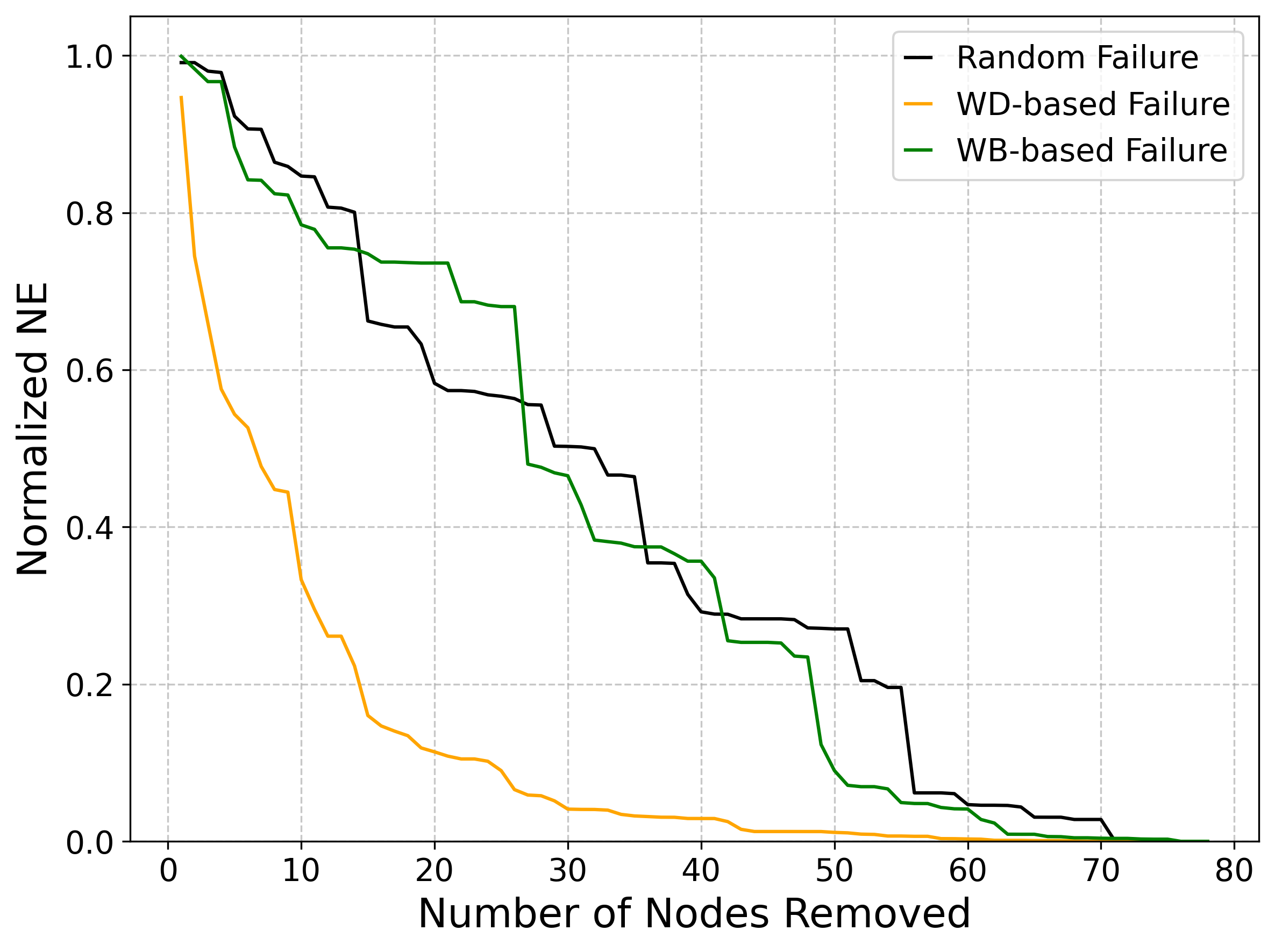}
   \caption{\textbf{Failure normalized weighted network efficiency (NE) under different node failure strategies}}
    \label{fig:failure}
\end{figure}

\subsection{Partial Node Disruptions}
In contrast to complete node removals, real-world disruptions often result in partial loss of functionality, such as degraded performance, congestion, or intermittent failures. \textbf{Figure~\ref{fig:partial}} illustrates the intermodal network's response to progressive node degradation at three levels (100\% for complete failure, 60\%, and 30\%), under three disruption strategies: Random, Weighted Degree (WD), and Weighted Betweenness (WB). Functional performance is measured using Normalized Network Efficiency (NE). The results show a clear relationship between degradation intensity, targeting strategy, and system-wide performance. As expected, greater degradation leads to steeper NE decline, but even partial failures significantly affect functionality. Under 60\% and 30\% degradation, NE falls below 20\% and 60\%, respectively. Nonetheless, the network retains relatively high performance, over 80\% NE with 30 degraded nodes at 30\% failure\rev{,} particularly under random failures, indicating resilience to non-targeted disruptions.
\begin{figure}[!htbp]
  \centering
  \includegraphics[width=\textwidth]{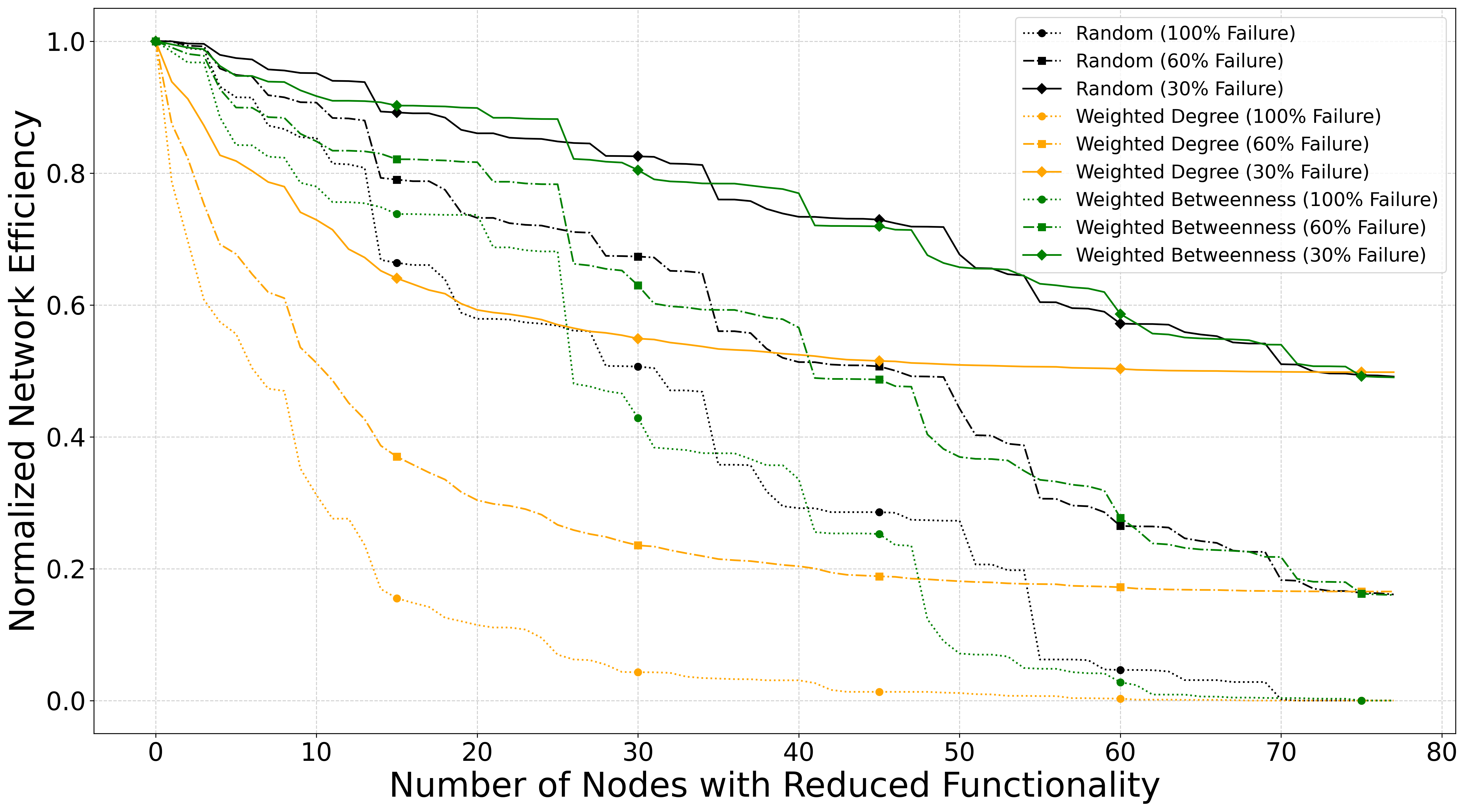}
    \caption{\textbf{Impact of partial node functionality on normalized network efficiency under different disruption strategies}}
    \label{fig:partial}

\end{figure}

A notable finding is that 60\% degradation of WD nodes causes a sharper NE decline, up to approximately 45 nodes, than 100\% failure of WB or Random nodes. \rev{This outcome is fundamentally driven by structural differences in how centrality metrics identify critical nodes. High-WD nodes are structurally characterized by their high connectivity to freight-demand-intensive links, resulting in strong concentration of total network flow. Consequently, partial degradation directly scales down the largest demand-weighted contributions to NE. Since NE is computed as a demand-weighted sum normalized by shortest-path distance, reducing flows at these nodes produces an immediate and disproportionate decline, even though network topology and path distances remain unchanged. In contrast, high-WB nodes are identified by their position along shortest paths rather than by the magnitude of freight they handle, so their removal initially affects fewer high-tonnage OD pairs. To further illustrate this structural distinction, Table \ref{tab:flow_wdc_wbc} compares the top-ranked nodes under WDC-based partial degradation and WBC-based full removal. The limited overlap between the two rankings highlights that flow-weighted and path-based centrality metrics identify fundamentally different critical structures within the same network, reinforcing the importance of metric choice in resilience analysis.} WD-based disruptions are consistently the most damaging across all failure levels, with NE dropping below 35\% after just 20 degraded nodes at 60\% functionality. However, their impact plateaus after approximately 50 degraded nodes, \rev{indicating that WD-based targeting rapidly degrades the limited subset of nodes that concentrate the majority of freight demand, after which further degradations predominantly affect nodes with substantially smaller demand-weighted contributions to NE.}
\rev{In contrast,} WB-based degradation produces a more gradual but sustained decline in NE, reflecting \rev{the more dispersed structural nature of betweenness centrality, in which many nodes contribute incrementally to shortest-path structure rather than dominating freight demand, so successive removals progressively reduce network efficiency through cumulative structural effects.} An additional insight arises from the intertwined NE trajectories of WB and Random disruptions. Multiple crossover points indicate that neither strategy is consistently more damaging. \rev{Their relative impact depends on both the number and identity of affected nodes, as well as the specific \rev{OD} flows they support. These overlapping and intersecting trajectories illustrate the heterogeneous importance of nodes within each disruption strategy and highlight the nonlinear, context-dependent nature of functional resilience in intermodal freight networks.}

\begin{table}[]
\centering
\caption{\rev{Comparison of top-ranked nodes under partial degradation: Remaining Network Flow Capacity under WDC (60\% degradation) and WBC (100\% degradation)}}
\label{tab:flow_wdc_wbc}
\rev{
\begin{tabular}{|c|cc|cc|}

\hline
              & \multicolumn{2}{c|}{\textbf{WDC (60\% degradation)}}                                                                            & \multicolumn{2}{c|}{\textbf{WBC (100\% degradation)}}                                                                           \\ \hline
\textbf{Rank} & \multicolumn{1}{c|}{\textbf{Node ID}} & \textbf{\begin{tabular}[c]{@{}c@{}}Remaining Network \\ Capacity (Tonnes)\end{tabular}} & \multicolumn{1}{c|}{\textbf{Node ID}} & \textbf{\begin{tabular}[c]{@{}c@{}}Remaining Network \\ Capacity (Tonnes)\end{tabular}} \\ \hline
1             & \multicolumn{1}{c|}{105}              & 198,464.32                                                                              & \multicolumn{1}{c|}{151}              & 213,499.84                                                                              \\ \hline
2             & \multicolumn{1}{c|}{230}              & 187,900.32                                                                              & \multicolumn{1}{c|}{154}              & 208,067.95                                                                              \\ \hline
3             & \multicolumn{1}{c|}{204}              & 174,745.72                                                                              & \multicolumn{1}{c|}{148}              & 204,108.29                                                                              \\ \hline
4             & \multicolumn{1}{c|}{241}              & 161,958.11                                                                              & \multicolumn{1}{c|}{149}              & 204,108.29                                                                              \\ \hline
5             & \multicolumn{1}{c|}{205}              & 157,085.66                                                                              & \multicolumn{1}{c|}{204}              & 183,685.87                                                                              \\ \hline
6             & \multicolumn{1}{c|}{217}              & 149,194.15                                                                              & \multicolumn{1}{c|}{241}              & 166,678.62                                                                              \\ \hline
7             & \multicolumn{1}{c|}{140}              & 140,965.30                                                                              & \multicolumn{1}{c|}{201}              & 166,215.43                                                                              \\ \hline
8             & \multicolumn{1}{c|}{138}              & 135,574.49                                                                              & \multicolumn{1}{c|}{205}              & 163,513.28                                                                              \\ \hline
9             & \multicolumn{1}{c|}{137}              & 131,597.52                                                                              & \multicolumn{1}{c|}{208}              & 162,249.34                                                                              \\ \hline
10            & \multicolumn{1}{c|}{212}              & 123,785.83                                                                              & \multicolumn{1}{c|}{133}              & 150,367.02                                                                              \\ \hline

\end{tabular}
}
\end{table}

The findings underscore the nonlinear relationship between node degradation and system-wide performance, where even moderate efficiency losses at key flow-critical nodes can trigger cascading disruptions. From a resilience planning perspective, these findings reinforce the need to strengthen or partially protect flow-critical nodes, particularly those with high flow-weighted degree centrality. Prioritizing these nodes can significantly slow the rate of functional decline, especially in scenarios such as labor disruptions, cyberattacks, or equipment failures, where complete node loss is unlikely but partial degradation is common. 

\subsection{Limitations}
\rev{Several limitations merit consideration: freight demand is based on FAF projections with national modal scaling, the efficiency metric does not capture capacity constraints, congestion, or recovery, and disruptions are modeled deterministically. Future research could address these limitations by incorporating capacity-aware routing, stochastic failures, and recovery dynamics.}

\section{Conclusion}

This study evaluated the resilience of the U.S. intermodal freight transportation network through a combination of structural and flow-based analyses. Results show that the network is highly robust to random failures but vulnerable to targeted disruptions, particularly those based on recalculated betweenness centrality. Node removals had a more severe impact than edge removals, underscoring the critical role of intermodal hubs. Flow-weighted metrics revealed that even partial degradation of high-volume nodes can significantly reduce system efficiency. \rev{When such flow-critical nodes are disrupted, freight is often rerouted through longer or less efficient routes and alternative modes, which can increase fuel consumption and associated emissions, directly supporting the RECOIL objective of quantifying disruption-induced inefficiencies. From a practical perspective, the results indicate that resilience planning should extend beyond structural connectivity to account for freight flow concentration and partial disruptions. Moreover, widespread or prolonged failures at critical hubs may propagate beyond the transportation network, potentially affecting regional supply chains and local economic activity. The proposed framework enables planners and operators to stress-test intermodal networks, identify flow-critical hubs, and prioritize mitigation strategies under realistic degradation scenarios.} Overall, the findings highlight the importance of identifying structurally and functionally critical nodes to guide infrastructure protection planning. The proposed framework offers a practical tool for enhancing freight system resilience against diverse and evolving threats.

\textbf{Acknowledgment}
\label{acknowledgement}

This research was conducted at the University of Tennessee, Knoxville, in cooperation with the Oak Ridge National Laboratory. This work was supported in part by the \rev{U.S.} Department of Energy’s Advanced Research Projects Agency-Energy (ARPA-E) under the project (\#DE-AR0001780), titled ``A Cognitive Freight Transportation Digital Twin for Resiliency and Emission Control Through Optimizing Intermodal Logistics'' (RECOIL). This manuscript has been authored by UT-Battelle, LLC, under contract DE-AC05-00OR22725 with the  \rev{U.S.}  Department of Energy (DOE). The  \rev{U.S.}  government retains and the publisher, by accepting the article for publication, acknowledges that the  \rev{U.S.}  government retains a nonexclusive, paid-up, irrevocable, worldwide license to publish or reproduce the published form of this manuscript, or allow others to do so, for  \rev{U.S.}  government purposes. DOE will provide public access to these results of federally sponsored research in accordance with the DOE Public Access Plan (http://energy.gov/downloads/doe-public-access-plan).

We used ChatGPT (version 5.0), a large language model developed by OpenAI, to assist in minor language editing and grammar checking during manuscript preparation. All substantive content, analysis, and interpretations were conceived, written, and verified by the authors. The authors have carefully reviewed and confirmed the accuracy and validity of all content and citations, and accept full responsibility for the entirety of the manuscript.

\textbf{Author Contributions}
\label{author_contributions}

The authors confirm their contribution to the paper as follows: \textit{Study Conception and Design:} all authors (Aliza Sharmin, Bharat Sharma, Mustafa Can Camur, Olufemi A. Omitaomu, Xueping Li); \textit{Data Collection:} all authors; \textit{Analysis and Interpretation of Results:} Aliza Sharmin, Bharat Sharma, Mustafa Can Camur; \textit{Draft Manuscript Preparation:} all authors. All authors review the results and approve the final version of the manuscript.

\textbf{Conflict of Interest}
The authors declare that they have no potential conflicts of interest with respect to the research, authorship, and/or publication of this article.

\bibliographystyle{unsrt}  
\bibliography{trb_template}  






\end{document}